\documentstyle[preprint,aps]{revtex}

\begin{document}

\tighten

\draft

\preprint{\today}

\title{Nonlinear resonant tunnelling through double-barrier
structures}

\author{Enrique Diez$^{\dag}$, Francisco Dom\'{\i}nguez-Adame$^{\ddag}$,
and Angel S\'{a}nchez$^{\dag}$}

\address{$^{\dag}$Escuela Polit\'{e}cnica Superior, Universidad
Carlos III de Madrid, c./ Butarque 15 \\ E-28911 Legan\'{e}s,
Madrid, Spain\\
$^{\ddag}$Departamento de F\'{\i}sica de Materiales,
Facultad de F\'{\i}sicas, Universidad Complutense,\\
E-28040 Madrid, Spain}

\maketitle

\begin{abstract}

We study resonant tunnelling through double-barrier structures under an
applied bias voltage, in which nonlinearities due to self-interaction of
electrons in the barrier regions are included.  As an approximation, we
concern ourselves with thin barriers simulated by $\delta$-function
potentials.  This approximation allows for an analytical expression of
the transmission probability through the structure.  We show that the
typical peaks due to resonant tunneling decrease and broaden as
nonlinearity increases.  The main conclusion is that nonlinear effects
degrade the peak-to-valley ratio but improve the maximum operation
frequency of the resonant tunnelling devices.

\end{abstract}


\narrowtext

\section{Introduction}

Resonant tunnelling (RT) through a heterostructure quantum well has
recently attracted considerable attention because of its applications as
ultra-high-speed electronic devices.  As an example, RT in GaAs/GaAlAs
double-barriers at THz frecuencies has been reported in the literature
\cite{Sollner}.  The way RT arises is a basic phenomenom which can be
explained with elementary quantum mechanics \cite{AJP}: there exists a
dramatic increase of the transmission probability whenever the energy of
an incident electron matches one of the unoccupied quasi-bound states
inside the well.  The energy of the incident electron is close to the
Fermi energy, which is constant in the device.  Therefore, to obtain a
RT current in practice, one must modify the energy of the discrete level
in the well applying a bias voltage.  If initially there are no
electrons in the well and the electrons are forced to tunnel by the
applied voltage, after a time of a few $\sim \hbar /\Gamma$, $\Gamma$
being the width of the discrete level, high transmission is achieved by
multiple reflection.  This mechanism is physically analogous to what
occurs in a Fabry-Perot resonator.  Since multiple scattering of waves
is required for RT, it can be realized that such a phenomenon depends
crucially on the {\em linear} superposition principle.  Then a natural
question arises in this context, namely how spatial nonlinearities
affect RT. Nonlinearity may arise in semiconductors when electrons
polarize the surrounding medium, which reacts to change the electron
state by a fedback process, or when electron-electron interaction is
taken into account.  There are two important magnitudes which, in
principle, could be largely modified by nonlinearity: the peak-to-valley
ratio and the level width $\Gamma$, which is related to the reliability
of the device at high frequencies.  For instance, several satellite
peaks in addition to the main peak have recently been found via coupled
electron-phonon states in RT double-barrier heterostructures
\cite{Figielski}. Moreover, the occurrence of nonlinearities opens the
possibility of studying new and interesting phenomena in semiconductor
devices, like bistability or chaotic wavefunctions \cite{Hawrylak}

In this Letter we focus our attention on a simple model to study
nonlinear effects in double-barrier heterostructures.  We consider very
narrow barriers of a nonlinear material embedded in a linear material.
We simulate the two barriers by means of $\delta$-function potentials,
which allow for an analytical treatment even in the presence of
nonlinearities.  Similar models have been proposed to study wave
propagation in nonlinear periodic media \cite{Grabowski} and nonlinear
Stark-Wannier resonances \cite{Cota}.  Thus, the propagation of a
stationary wave in the nonlinear double-barrier heterostructure is
described by means of the following Schr\"odinger equation (we use units
such that $\hbar=m=1$)
\begin{equation}
-\Psi''(x)+[\delta(x)+\delta(x-a)]\,[ \beta +\tilde{\alpha} \Psi(x)
\Psi^*(x)]\,\Psi(x)+V(x)\Psi(x)=E\>\Psi(x).
\label{Sch}
\end{equation}
The parameter $\beta$ is the strength of the $\delta$-potential in the
linear case (i.\ e.\ the area of the semiconductor barrier) and
$\tilde{\alpha}$ measures the strength of the nonlinear coupling.  The
width of the quantum well is denoted by $a$. The potential due to
the applied bias is $V(x)$.  This potential may be simulated by step
functions, which have been demonstrated to be a good approach to
linearly rising potentials, reproducing accurately the properties of the
electronic states, even the subtle ones \cite{Bentosela}.  Therefore we
take
\begin{equation}
V(x)=\left\{ \begin{array}{cl} 0, & x<0, \\
          -\,{\displaystyle V\over \displaystyle 2}, & 0<x<a, \\
                             -V,  & x>a,
             \end{array} \right.
\label{potentia}
\end{equation}
where $V$ is the applied voltage.

The solution of Eq.~(\ref{Sch}) in the linear semiconductor is a
combination of travelling waves. As usual in scattering problems, we
assume an electron incident from the left and evaluate the transmission
and reflection probabilities when passing through the structure. To
this end we take
\begin{equation}
\Psi(x)=\left\{\begin{array}{ll}A\left(e^{ik_0x}+re^{-ik_0x}\right),
& x<0, \\
A\left(Ce^{ik_1x}+De^{-ik_1x}\right) & 0<x<a, \\
Ate^{ik_2x} & x>a. \end{array} \right.
\label{wavefunction}
\end{equation}
$t$ and $r$ denote the transmission and reflection amplitudes, and we
have defined for simplicity $k_0^2=E$, $k_1^2=E+V/2$, and $k_2^2=E+V$.
The transmission probability is then found to be $T=(k_0/k_2)|t|^2$.  To
compute it we must set appropriate boundary conditions at the barriers,
in order to eliminate the two unknowns $C$ and $D$.  These can be
obtained by integrating Eq.~(\ref{Sch}) around $x=0$ and $x=a$.  In so
doing, we find that the wave function itself is continuous whereas its
derivative satisfies
\begin{eqnarray}
\Psi'(0^+)-\Psi'(0^-)=\left[ \beta +\tilde{\alpha} |\Psi(0)|^2 \right]\,
\Psi(0),\nonumber \\
\Psi'(a^+)-\Psi'(a^-)=\left[ \beta +\tilde{\alpha} |\Psi(a)|^2 \right]\,
\Psi(a) \label{boudary}
\end{eqnarray}
Using these boundary conditions and solution (\ref{wavefunction}) we
have after a little algebra
\begin{mathletters}
\label{final}
\begin{eqnarray}
r&=&t\left\{ \cos k_1a + {k_1\over \sin k_1a} [ (\beta + \alpha
|t|^2)-ik_2] \right\}-1, \label{finala}\\
t&=&(1+r){\sin k_1a\over k_1} \left\{ \left[ k_1 \cot k_1 a+ik_0+\beta
+\alpha |1+r|^2\right]-2ik_0r\right\}. \label{finalb}
\end{eqnarray}
\end{mathletters}
where $\alpha\equiv \tilde{\alpha}|A|^2$.  Finally, inserting
(\ref{finala}) in(\ref{finalb}) we obtain an expresion for the
transmission amplitude, and thus the transmission probability can be
found for a given set of parameters ($E$, $\alpha$, $\beta$, $a$ and
$V$).

Figure~\ref{fig1} shows the transmission probability as a function of
the applied voltage $V$ and the nonlinear coupling $\alpha$.  The energy
of the particle is taken to be $E=2$, $\beta=10$, and $a=1.5$.  It is
important to mention that the qualitative aspects of the results remain
almost unchanged when varying the parameters $E$, $\beta$, and $a$.
 From this plot it is clear that several peaks appear on incresing the
applied voltage, no matter the value of the nonlinear coupling $\alpha$.
Each peak corresponds to a quasi-bound state level in the well.  It is
also apparent that the width of these peaks is relatively large,
specially for higher order ones.  At this point we should remark that
the width of the peaks appearing in Fig.~\ref{fig1} (transmission
probability {\em versus\/} applied voltage) is not exactly equal to
$\Gamma$, i.\ e., the level width.  Nevertheless, assuming that the
lifetime is not strongly dependent on the position of the level (so that
it remains almost unchanged under minor variations of the applied
voltage), it is clear that the width of the resonance peak in the
transmission coefficient is very close to $\Gamma$.  Hence we can
accurately determine $\Gamma$ from Fig.~\ref{fig1}.  The fact that
$\Gamma$ is relatively large is easily understood by considering that
the barriers are, in fact, very narrow, so that the coupling of the
quantum well states with the continuum is very strong.  Thus the
lifetime is very short and, accordingly, $\Gamma$ is large.  Also
$\Gamma$ increases upon increasing the order of the quasi-bound state in
the well, as expected.  Now let us comment the main effects of
nonlinearity.  In Fig.~\ref{fig1} one can see that the peak-to-valley
ratio decreases dramatically on increasing $\alpha$ from $-30$ up to
$30$.  Moreover, it is also clear that the resonances shift to higher
voltages when increasing $\alpha>0$ but, on the contrary, resonances
shift to lower voltages for $\alpha<0$.  This means that the quasi-bound
states within the well region are raised on incresing nonlinearity.
With more elaborated models, it has been recently found a similar shift
to higher voltages whenever Coulomb repulsion between electrons
increases in quasi-one-dimensional systems \cite{Nono}.  Hence we are
led to the conclusion that our model retains most of the physics of
resonant tunneling, since Coulomb repulsion can be regarded as a
possible origin of nonlinearities within the one-particle approach.

Since the width of the peaks is a very important parameter for the
design of high-speed devices, we have also studied $\Gamma$ as a
function of the nonlinear parameter.  Results are shown in
Fig.~\ref{fig2} for three different values of the nonlinear parameter
$\alpha$.  Besides the relative shift of the peaks above mentioned, it
is seen that $\Gamma$ decreases with $\alpha$ for peaks of the same
order.  Hence we are led to the conclusion that, in order to improve the
quality of the RT device, one should look for a compromise between the
peak-to-valley ratio which decreases with $\alpha$, and the width of the
levels which also decrease.  Hence, on increasing nonlinearity, one
obtains a lower RT current but, in principle, a higher operation
frequency.

A possible extension of this work concerns narrow-gap semiconductors for
RT \cite{Gill,Beresford}.  In this case RT takes place by interband
tunnelling mechanism, and a simple one-band description is not longer
valid.  Instead, one must use more elaborated band structure including
the effects of several bands of the host semiconductors.  One of the
simplest model is the two-band model \cite{Callaway}, in which
envelope-functions of the conduction- and valence-bands are included.
The equation of motion is then analogous to the Dirac equation for
relativistic electrons.  Hence, Eq.~\ref{Sch} should be replaced by a
Dirac-like equation with nonlinear terms.  Fortunately, the nonlinear
Dirac equation with $\delta$-function potentials is also exactly
solvable\cite{JPA} and, therefore, the generalization of the present
work to narrow-gap heterostructures seems to be achievable.  In
addition, more realistic and elaborated models are indeed required to
describe nonlinear electron dynamics through thick barriers even if only
one envelope function is used, for which the $\delta$-function approach
is no longer valid.  Work in that direction is in progress.


F.\ D-A.\ acknowledges support from UCM through project PR161/93-4811.
A.\ S.\ acknowledges partial support from C.I.C.\ y T.\ (Spain) through
project PB92-0248 and by the European Union Human Capital and Mobility
Programme through contract ERBCHRXCT930413.

\begin{figure}
\caption{Transmission probability $T$ as a function of the applied
voltage $V$ and the nonlinear coupling $\alpha$ for $E=2$, $\beta=10$,
and $a=1.5$.}
\label{fig1}
\end{figure}

\begin{figure}
\caption{Transmission probability $T$ as a function of the applied
voltage $V$ for $E=2$, $\beta=10$, $a=1.5$, and (a) $\alpha=-10$, (b)
$0$, and (c) $+10$.}
\label{fig2}
\end{figure}


\begin{references}

\bibitem{Sollner} T.\ C.\ L.\ G.\ Sollner, W.\ D.\ Goodhue, P.\ E.\
Tannenwald, C.\ D.\ Parker, and D.\ D.\ Peck, Appl.\ Phys.\ Lett.\ {\bf
43}, 588 (1984).

\bibitem{AJP} B.\ M\'endez and F.\ Dom\'{\i}nguez-Adame, Am.\ J.\ Phys.\
{\bf 62}, 143 (1994).

\bibitem{Figielski} T.\ Figielski, A.\ M\c{a}kosa, T. Wosi\'nski, P.\
C.\ Harness and K.\ E.\ Singer, Solid Stat. Commun. {\bf 91}, 913
(1994).

\bibitem{Hawrylak} P.\ Hawrylak, M.\ Grabowski, and P.\ Wilson, Phys.\
Rev.\ B {\bf 40}, 6398 (1989).

\bibitem{Grabowski} M.\ Grabowski and P.\ Hawrylak, Phys.\ Rev.\ B {\bf
41}, 5783 (1990).

\bibitem{Cota} E.\ Cota, J.\ V.\ Jos\'e, and G.\ Monsivais, J.\ Phys.\
A:\ Math.\ Gen.\ {\bf 25}, L57 (1992).

\bibitem{Bentosela} F.\ Bentosela, V.\ Grecchi, and F.\ Zironi, J.\
Phys.\ C:\ Solid State Phys.\ {\bf 15}, 719 (1982).

\bibitem{Nono} S.\ Nonoyama, A.\ Oguri, Y.\ Asano, and S.\ Makeawa,
Phys.\ Rev.\ B {\bf 50}, 2667 (1994).

\bibitem{Gill} T.\ C.\ Gill and D.\ A.\ Collins, Semicond.\ Sci.\
Technol.\ {\bf 5}, S1 (1990).

\bibitem{Beresford} R.\ Beresford, L.\ F.\ Luo, and W.\ I.\ Wang,
Semicond.\ Sci.\ Technol.\ {\bf 5}, S195 (1990).

\bibitem{Callaway} J.\ Callaway, {\em Quantum Theory of the Solid
State}, (Academic Press, New York, 1991).

\bibitem{JPA} F.\ Dom\'{\i}nguez-Adame, J.\ Phys.\ A:\ Math.\ Gen.\ {\bf
26}, 3863 (1993).

\end{references}
\end{document}